\documentclass[prb,twocolumn,superscriptaddress,showpacs]{revtex4}

\usepackage[dvips]{graphicx}

\begin{document}

\title{Photoconductance of organic single-molecule contacts}

\author{J. K. Viljas}
\email{janne.viljas@kit.edu}
\affiliation{Institut f\"ur Theoretische Festk\"orperphysik 
and DFG-Center for Functional Nanostructures, 
Universit\"at Karlsruhe, D-76128 Karlsruhe, Germany}
\affiliation{Forschungszentrum Karlsruhe, 
Institut f\"ur Nanotechnologie, D-76021 Karlsruhe, Germany }

\author{F. Pauly}
\affiliation{Institut f\"ur Theoretische Festk\"orperphysik
and DFG-Center for Functional Nanostructures,  
Universit\"at Karlsruhe, D-76128 Karlsruhe, Germany}
\affiliation{Forschungszentrum Karlsruhe, 
Institut f\"ur Nanotechnologie, D-76021 Karlsruhe, Germany }

\author{J. C. Cuevas}
\affiliation{Departamento de F\'{\i}sica Te\'orica de la Materia
Condensada, Universidad Aut\'onoma de Madrid, E-28049 Madrid, Spain}
\affiliation{Institut f\"ur Theoretische Festk\"orperphysik
and DFG-Center for Functional Nanostructures, 
Universit\"at Karlsruhe, D-76128 Karlsruhe, Germany}
\affiliation{Forschungszentrum Karlsruhe, 
Institut f\"ur Nanotechnologie, D-76021 Karlsruhe, Germany }

\date{\today}

\begin{abstract}
  We study the dc conductance of organic single-molecule contacts in
  the presence of external electromagnetic radiation
  (photoconductance).  In agreement with previous predictions, we find
  that the radiation can lead to large enhancements of the conductance
  of such contacts by bringing off-resonant levels into resonance
  through photoassisted processes.  In our analysis we make use of the
  simplifying fact that, under certain assumptions, the
  photoconductance can be expressed in terms of the transmission
  function in the absence of the radiation. The conductance
  enhancement is demonstrated for oligophenylene molecules between
  gold electrodes, whose electronic structure is calculated based on
  density-functional theory.  It is shown that the exponential decay
  of the conductance with the length of the molecule can be replaced
  by a length-independent value in the presence of radiation.
\end{abstract}



\maketitle
 
The use of electromagnetic radiation, such as laser light, can provide
a convenient ``handle'' for controlling the conductance of atomic and
single-molecule contacts.\cite{Lee05} It has, for example, been
demonstrated that by applying light of a certain frequency, some
photochromic molecules can be made to change their conformation even
when contacted to metallic electrodes.\cite{Dulic03,He05} Devices
based on such molecules could act as molecular optoelectronic
switches.\cite{Li04,Zhang04,Zhuang05,Huang06} According to theoretical
predictions, it may also be possible to construct molecular electronic
devices based on other radiation-induced
phenomena,\cite{Tikhonov02,Lehmann02,Urdaneta04,Welack06,Bittner05,Baer04}
and several groups are currently working towards experiments of this
type.\cite{Guhr06,Meyer07,Simmons07}

One of the first things to be addressed in interpreting conductance
measurements on atomic-sized contacts exposed to external radiation is
the role played by heating effects, such as electronic excitations and
thermal expansion.\cite{Guhr06,Grafstrom02} Another phenomenon to be
considered is the excitation of local plasmonic modes and related
field-enhancement effects.\cite{Grafstrom02,Sanders06} Due to the
complexity of the problem, no comprehensive theory exists at the
moment. However, we have recently put forward a description based on
photoassisted transport.\cite{Viljas07} We found that, depending on
the metal and the radiation frequency, the effect of irradiation on
atomic contacts can be either an increase or a decrease in the
conductance. We also demonstrated that the approximate behavior of the
photoconductance can be predicted based on the transmission function $T(E)$
in the absence of the radiation, without resorting to complicated
numerical simulations of the ac-driven transport.

In this brief report we discuss the possible outcome of experiments
with laser-irradiated organic single-molecule contacts between two
metallic electrodes. We shall argue that for junctions where the Fermi
energy of the metal lies in the gap between the highest occupied
(HOMO) and lowest unoccupied (LUMO) molecular orbital, photoassisted
processes can lead to enhancements of the dc conductance by orders of
magnitude. This conclusion has already been made
previously,\cite{Tikhonov02} but here we want to describe the origin
of the effect based on the properties of $T(E)$, along the lines of
Ref.\ \onlinecite{Viljas07}.  We demonstrate the effect for
oligophenylene contacts with varying numbers of phenyl rings, whose
electronic structure is calculated using density-functional theory.
While in the absence of radiation the conductance decays
exponentially, in the presence of radiation it may become almost
independent of the length of the molecule.  This effect should become
important already for light frequencies lower than those needed for
internal transitions between the HOMO and LUMO
levels.\cite{Galperin05}

In a typical experiment with laser-irradiated atomic-sized contacts,
the laser spot diameters are on the order of micrometers.\cite{Guhr06}
It is therefore much more likely that an incoming photon interacts
with the metallic electrode than with the molecule itself. However, at
frequencies below the plasma frequency and the electronic interband
transition threshold, most of the incoming light is reflected.  This
is the result of a collective screening response of the electrons.  If
the light is polarized along the axis of the contact, this generates
an oscillating voltage over the contact at the frequency $\omega$ of
the radiation. The photoassisted transport through the contact can
then be described in terms of ``sidebands'', where the Fermi-level
electrons appear to approach the contact at energies shifted from the
equilibrium Fermi level by an integer multiple of the photon energy
$\hbar\omega$ (see Refs.\
\onlinecite{Tien63,Tikhonov02,Pedersen98,Moskalets04,Kohler04,Platero04,Viljas07,Tucker85}).
In this view, the role of the molecule is, most importantly, to
provide the transmission landscape according to which the incoming
electrons will be transmitted or reflected. Thus the photoassisted
transport can be seen as probing the transmission function at energies
away from the Fermi level.

Assuming a symmetric junction, low temperature, a double-step voltage
profile (i.e. vanishing ac electric field on the molecule), as well as
``wide-band'' leads, the zero-bias dc conductance in the presence of
external radiation of frequency $\omega$ (i.e. the photoconductance)
can be described by the expression\cite{Viljas07}
\begin{equation}\label{e.g}
G_{dc}(\omega)=G_0\sum_{l=-\infty}^{\infty}
[J_l(\alpha/2)]^2T(E_F+l\hbar\omega).
\end{equation}
Here $G_0=2e^2/h$ is the quantum of conductance, $E_F$ is the Fermi
energy, $l$ is the sideband index, and $J_l(x)$ are Bessel functions
of the first kind. Their argument involves
$\alpha=eV_{ac}/\hbar\omega$, where $V_{ac}$ is the amplitude of the
induced ac voltage. The latter is unknown in practice, because it
depends on the environment of the junction, the polarization of the
light,\cite{Note1} and the frequency itself.  Especially if $\omega$
happens to be in resonance with local plasmonic excitations, $V_{ac}$
can be strongly enhanced. For our purposes it is sufficient to treat
$\alpha$ simply as a parameter.  In the metallic atomic contacts
described in Ref.\ \onlinecite{Viljas07} the transmission functions
are rather flat, and so the changes in conductance typically remain on
the order of a few percent for reasonable ac amplitudes. In contrast,
molecular junctions often exhibit large gaps in $T(E)$, corresponding
roughly to the energy region between the HOMO and LUMO orbitals of the
isolated molecule. The Fermi energy lies somewhere in the gap and thus
the dc conductance of the junction in the absence of radiation is very
low. It is then to be expected from Eq.\ (\ref{e.g}) that an external
frequency corresponding to the smaller of the energy differences
between $E_F$ and the two gap edges can lead to a considerable
enhancement of the conductance. This is because the photoassisted
processes essentially change the character of the transport from
off-resonant to resonant tunneling.\cite{Tikhonov02}

\begin{figure}[!tb]
\includegraphics[width=0.85\linewidth,clip=]{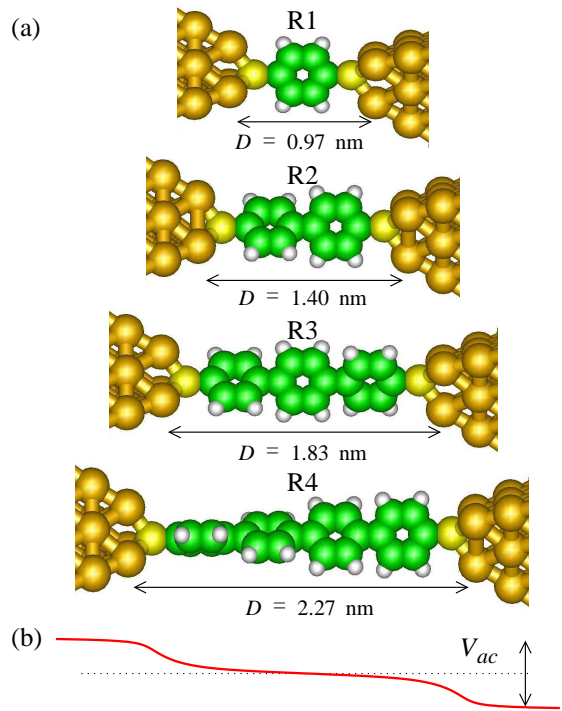}
\caption{(Color online) (a) The four molecular contacts R1--R4 we have 
studied, containing oligophenylenes with one to four phenyl rings and
coupled to Au [111] pyramids through sulfur atoms. (b) Our model
assumes the induced ac voltage $V_{ac}$ to drop in a double-step
manner, as illustrated here schematically for contact R4. 
}
\label{f.contacts}
\end{figure}

The photoassisted conductance-enhancement described above should be
contrasted with the light-induced (rectification) currents studied in
Ref.\ \onlinecite{Galperin05}, for example. The latter are due to a
direct internal pumping of electrons between the HOMO and LUMO levels
of a molecule coupled weakly and asymmetrically to the electrodes.
Such internal transitions may become dominant for frequencies
exceeding the HOMO-LUMO gap. In contrast, the photoassisted processes
described by Eq.\ (\ref{e.g}) involve photon emission and absorption
in the contact regions between the molecule and the electrodes, while
the transport on the molecule is assumed to be elastic. These
processes should set in already for $\hbar\omega$ smaller than half of
the gap. In the following we consider low enough frequencies, such
that the internal electronic transitions can be assumed to be
unimportant. On the other hand, in order to apply Eq.\ (\ref{e.g}), we
must consider (approximately) symmetric contacts. Therefore,
significant rectification effects will be absent in any case.

To illustrate the above ideas for realistic molecules, we have used
density-functional theory (DFT) to describe oligophenylene molecules
with varying numbers of phenyl rings bridging two gold electrodes.
The description of the electronic structure and the relaxation of the
geometries is done using the \textsc{Turbomole} quantum chemistry
package,\cite{Ahlrichs89} and the transmission functions are
calculated using Green's function
techniques.\cite{PaulyThesis,Wohlthat07} The four molecular junctions
considered here are shown in Fig.\ \ref{f.contacts}(a). They are
formed of oligophenylenes containing one to four phenyl rings and are
bonded to the fcc [111] gold pyramids through sulfur atoms. We refer
to them as R1 to R4, according to the number of rings. The
corresponding isolated molecules have HOMO-LUMO gaps of 3 eV or more,
and thus to induce internal transitions of electrons between the HOMO
and LUMO levels would require energies corresponding to blue or
ultraviolet light. On the other hand, the photoassisted effects we are
describing are expected to take effect already in the red or infrared
part of the spectrum. Note that the double-step ac voltage profile
assumed by Eq.\ (\ref{e.g}) is not unreasonable, since one can expect
the voltage to drop abruptly at the sulfur atoms due to partial
screening of the electric field on the
molecule.\cite{Datta97,Lang00,Mujica00,Damle01,Nitzan02,Baer04} See
Fig.\ \ref{f.contacts}(b) for an illustration.

\begin{figure}[!tb]
\includegraphics[width=0.9\linewidth,clip=]{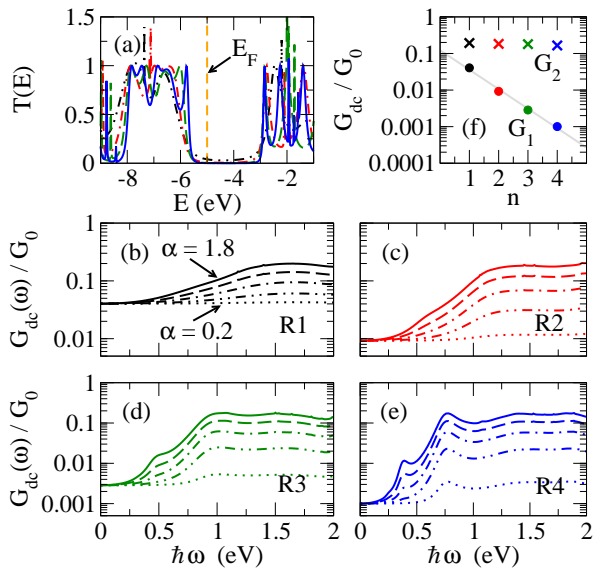}
\caption{(Color online) (a) Transmission versus energy [$T(E)$]
for the contacts R1--R4 in Fig.\ \ref{f.contacts} (dash-dot-dotted,
dash-dotted, dashed, and solid lines, respectively). 
(b)--(e) The photoconductance versus external frequency $\omega$ for
the contacts R1--R4, respectively. For each case the results for the
following values of $\alpha$ are shown: 0.2, 0.6, 1.0., 1.4, and 1.8,
in order of increasing conductance.  (f) The dc conductances in the
absence ($G_1$, dots) and presence ($G_2$, crosses) of radiation with
$\hbar\omega=1.5$ eV and $\alpha=1.8$ for an increasing number $n$ of
phenyl rings. The gray line is a fit of the $G_1$ results to an 
exponential law (see text).
}
\label{f.curves}
\end{figure}

Figure \ref{f.curves}(a) shows the $T(E)$ functions for the four
molecular contacts R1 to R4. All of them feature a gap on the order of
the HOMO-LUMO gap of the isolated molecules. The Fermi energy is in
the gap, somewhat closer to the HOMO than the LUMO edge.  The panels
(b)-(e) present the full results for $G_{dc}(\omega)$ for several
representative values of $\alpha$, as obtained from Eq.\ (\ref{e.g}).
Note that the zero-frequency results reproduce the conductances in the
absence of radiation, i.e. $G_{dc}(\omega=0)= G_0T(E_F)$ independently
of $\alpha$. When $\hbar\omega$ is increased to values above the
energy difference between $E_F$ and the HOMO edge of the gap,
$G_{dc}(\omega)$ can increase by orders of magnitude as one of the
sidebands comes into resonance with the molecular energy levels
[$l=-1$ in Eq.\ (\ref{e.g})]. For $\alpha\gtrsim1$ also ``two-photon''
processes ($l=-2$) begin to contribute, such that the conductance
enhancement begins already at lower frequencies (see Ref.\
\onlinecite{Note2}).  In Fig.\ \ref{f.curves}(f) we present the
results for $G_{dc}(\omega)$ in the absence of radiation ($G_1$, dots)
and in the presence of radiation with $\hbar\omega=1.5$ eV and
$\alpha=1.8$ ($G_2$, crosses) for an increasing number $n$ of phenyl
rings.  The value of $G_1$ exhibits an exponential decay,
characteristic of off-resonant tunneling.  The gray line represents a
fit to $G_1\propto G_0\exp(-\beta D)$, where $D$ is the distance
between the gold tips (see Fig.\ \ref{f.contacts}). We obtain an
attenuation factor of $\beta=2.8$ nm$^{-1}$, in agreement with
previous experimental and theoretical results.  \cite{Wold02,Kondo04}
In contrast, $G_2$ is almost independent of $n$, because the
photoassisted processes change the character of the transport to
resonant tunneling. Thus the conductance enhancement due to radiation 
is bigger for larger $n$.

We have also performed calculations with oligophenylene contacts,
where the conjugation of the molecules has been broken by side groups.
$T(E)$ then exhibits sharp resonances, leading to large fluctuations
in $G_{dc}(\omega)$. The maximal conductance enhancements can,
however, be even much larger than those in Fig.\ (\ref{f.curves}).

It must be emphasized that the calculations of the photoconductance
are based on ground-state DFT and Eq.\ (\ref{e.g}), which introduce
several simplifying assumptions. First, a proper treatment of the
electronic structure in the presence of time-dependent fields should
be based on more advanced techniques.\cite{Baer04,Kurth05} Second, the
ac-driven transport cannot in general be described in terms of independent
sidebands.\cite{Tikhonov02,Pedersen98,Moskalets04,Kohler04,Arrachea06}
Nevertheless, detailed tight-binding calculations\cite{Viljas07} for
atomic contacts reproduce the essential features of results obtained
with Eq.\ (\ref{e.g}).  In particular, the shape of the assumed ac
voltage profile [see Fig.\ \ref{f.contacts}(b)] does not seem to be
crucial.  As an additional check, we have studied chain models
describing contacts with gaps in $T(E)$ (not shown). In these
simulations, a linear ramp-like ac voltage profile (a constant ac
electric field) tends to reduce the conductance enhancement, but it
still remains an order-of-magnitude effect. Note that such a potential
profile was also assumed in Ref.\ \onlinecite{Tikhonov02}.  Indeed, in
the case of molecular junctions, the light-induced increase in the dc
conductance can be so huge that it is difficult to see how it could be
completely washed away in a more rigorous treatment.  The best chances
of measuring the effect in experiments would be at infrared
frequencies, such that the absorption of light and associated heating
effects are minimal.

Finally, it is worth noting that the photoconductance $G_{dc}(\omega)$
should not be confused with an ac conductance. The latter quantity has
also been discussed recently in the context of molecular
contacts.\cite{Baer04,Wu05} It is more difficult to describe
theoretically as well as to measure reliably, since the capacitance of
the junction and hence displacement currents will play a more
important role than in the case of the dc response.

In conclusion, we discussed the effect of external radiation on the
transport properties of organic single-molecule contacts between
metallic electrodes, where the Fermi energy lies in a gap of the
transmission function.  The importance of the collective response of
the leads to the radiation was emphasized, as compared with the
internal electronic transitions due to a direct pumping of the
molecule.  We have discussed how, under certain assumptions, the
photoconductance can be related to the transmission function of the
contact in the absence of the radiation. This relation was used to
analyze the radiation-induced conductance enhancement for
oligophenylene molecules of varying lengths in gold contacts.  It was
shown that the exponential decay of the dc conductance can be
replaced by a length-independent behavior as a result of the
photoassisted transport processes.

We acknowledge helpful discussions with Daniel Guhr, Elke Scheer, Paul
Leiderer, Marcelo Goffman, and Jan van Ruitenbeek.  The Quantum
Chemistry group of Reinhart Ahlrichs is thanked for providing us with
\textsc{Turbomole}.  This work was financially supported by the
Helmholtz Gemeinschaft (Contract No.\ VH-NG-029), by the DFG within
the CFN, and by the EU network BIMORE (Grant No. MRTN-CT-2006-035859).


\begin{thebibliography}{99}

\bibitem{Lee05}
 {T.-H.} Lee, {J. J.} Gonzalez, J. Zheng, and {R. M.} Dickson,
 %\emph{Single-Molecule Optoelectronics},
 Acc. Chem. Res. {\bf 38}, 534 (2005).

\bibitem{Dulic03}
 D. D\'ulic, {S. J.} {van der Molen}, T. Kudernac,
 {H. T.} Jonkman, {J. J. D.} {de Jong}, {T. N.} Bowden, 
 J. {van Esch}, {B. L.} Feringa, {B. J.} {van Wees},
 %\emph{One-Way Optoelectronic Switching of Photochromic 
 %Molecules on Gold},
 Phys. Rev. Lett. {\bf 91}, 207402 (2003);
 {S. J.} {van der Molen}, H. {van der Vegte}, T. Kudernac, I. Amin, 
 {B. L.} Feringa, and {B. J.} {van Wees},
 %\emph{Stochastic and photochromic switching of diarylethenes studied
 %by scanning tunneling microscopy},
 Nanotechnology {\bf 17}, 310 (2006);
 N. Katsonis, T. Kudernac, M. Walko, {S. J.} {van der Molen},
 {B. J.} {van Wees}, {B. L.} Feringa,
 %\emph{Reversible Conductance Switching of single Diarylethenes
 %on a Gold surface},
 Advanced Materials {\bf 18}, 1397 (2006).

\bibitem{He05}
 J. He, F. Chen, {P. A.} Liddell, J. Andr{\'e}asson, 
 {S. D.} Straight, D. Gust, {T. A.} Moore, {A. L.} Moore,
 J. Li, {O. F.} Sankey, {S. M.} Lindsay,
 %\emph{Switching of a photochromic molecule on gold electrodes:
 %single-molecule measurements},
 Nanotechnology {\bf 16}, 695 (2005).

\bibitem{Li04}
 J. Li, G. Speyer and {O. F.} Sankey,
 %\emph{J. Li and G. Speyer and O. F. Sankey},
 Phys. Rev. Lett. {\bf 93}, 248302 (2004).

\bibitem{Zhang04}
 C. Zhang, {M.-H.} Du, {H.-P.} Cheng, {X.-G.} Zhang, {A. E.} Roitberg
 and {J. L.} Krause, 
 %\emph{Coherent Electron Transport through an Azobenzene Molecule:
 %A Light-Driven Molecular Switch},
 Phys. Rev. Lett. {\bf 92}, 158301 (2004);
 C. Zhang, Y. He, {H.-P.} Cheng, Y. Xue, {M. A.} Ratner, {X.-G.} Zhang,
 and P. Krstic,
 %\emph{current-voltage characteristics through a single 
 %light-sensitive molecule},
 Phys. Rev. B {\bf 73}, 125445 (2006).
 
\bibitem{Zhuang05}
 M. Zhuang and M. Ernzerhof, 
 %\emph{Mechanism of a molecular electronic photoswitch},
 Phys. Rev. B {\bf 72}, 073104 (2005).

\bibitem{Huang06}
 J. Huang, Q. Li, H. Ren, H. Su, and J. Yang,
 %\emph{Single quintuple bond [PhCrCrPh] molecule as a possible 
 %molecular switch},
 J. Chem. Phys. {\bf 125}, 184713 (2006).

\bibitem{Tikhonov02}
 A. Tikhonov, {R. D.} Coalson, Y. Dahnovsky,
 %\emph{Calculating electron transport in a tight binding model of a 
 %field-driven molecular wire: Floquet theory approach},
 J. Chem. Phys. {\bf 116}, 10909 (2002);
 %\emph{Calculating electron current in a tight binding model of a 
 %field-driven molecular wire: Application to xylyl-dithiol},
 J. Chem. Phys. {\bf 117}, 567 (2002).

\bibitem{Lehmann02}
 J. Lehmann, S. Kohler, P. H\"anggi, and A. Nitzan, 
 %\emph{Molecular Wires Acting as Coherent Quantum Ratchets},
 Phys. Rev. Lett. {\bf 88}, 228305 (2002);
 J. Lehmann, S. Camalet, S. Kohler, and P. H\"anggi,
 %\emph{Laser controlled molecular switches and transistors},
 Chem. Phys. Lett. {\bf 368}, 282 (2003).

\bibitem{Urdaneta04}
 I. Urdaneta, A. Keller, O. Atabek, and V. Mujica, 
 %\emph{Laser-Assisted Conductance of Molecular Wires: 
 %Two-Photon Contributions},
 Int. J. Quant. Chem. {\bf 99}, 460 (2004);
 %\emph{A simple model for laser-electrode interaction and its role 
 %in photo-assisted electron transport processes in molecular interfaces}
 J. Phys. B: At. Mol. Opt. Phys. {\bf 38}, 3779 (2005).

\bibitem{Welack06}
 S. Welack, M. Schreiber, and U. Kleinekath\"ofer,
 %\emph{The influence of ultrafast laser pulses on electron transport
 %in molecular wires studies by non-Markovian density-matrix approach}
 J. Chem. Phys. {\bf 124}, 044712 (2006).

\bibitem{Bittner05}
 E. R. Bittner, S. Karabunarliev, and A. Ye,
 %\emph{Photoconductivity and current producing states in 
 %molecular semiconductors},
 J. Chem. Phys. {\bf 122}, 034707 (2005).

\bibitem{Baer04}
 R. Baer, T. Seideman, S. Ilani, and D. Neuhauser, 
 %\emph{Ab initio study of the alternating current impedance
 %of a molecular junction},
 J. Chem. Phys. {\bf 120}, 3387 (2004).

\bibitem{Guhr06}
 D. Guhr, D. Rettinger, J. Boneberg, A. Erbe, P. Leiderer,
 and E. Scheer,
 %\emph{Influence of chopped laser light onto the electronic
 %transport through atomic-sized contacts},
 %submitted to Journal of Microscopy
 cond-mat/0612117.

\bibitem{Meyer07}
 C. Meyer, {J. M.} Elzermann, and {L. P.} Kouwenhoven,
 %\emph{Photon-assisted Tunneling in a Carbon Nanotube
 %quantum Dot},
 Nano Lett. {\bf 7}, 295 (2007).

\bibitem{Simmons07}
 {J. M.} Simmons, I. In, {V. E.} Campbell, {T. J.} Mark, F. L\'eonard,
 P. Gopalan, and {M. A.} Eriksson,
 %\emph{Optically Modulated Conduction in Chromophore-Functionalized
 %Single-Wall Carbon Nanotubes},
 Phys. Rev. Lett. {\bf 98}, 086802 (2007).

\bibitem{Grafstrom02}
 S. Grafstr\"om,
 %\emph{Photoassisted scanning tunneling microscopy},
 J. Appl. Phys. {\bf 91}, 1717 (2002).

\bibitem{Sanders06}
 {A. W.} Sanders, {D. A.} Routenberg, {B. J.} Wiley, Y. Xia, 
 {E. R.} Dufresne, and {M. A.} Reed,
 %\emph{Observation of Plasmon Propagation, Redirection, and Fan-Out 
 %in Silver Nanowires}
 Nano Lett. {\bf 6}, 6, 1822 (2006).

\bibitem{Viljas07}
 {J. K.} Viljas and {J. C.} Cuevas
 %\emph{Role of electronic structure in photoassisted transport through
 %atomic-sized contacts},
 Phys. Rev. B {\bf 75}, 075406 (2007).

\bibitem{Galperin05}
 M. Galperin and A. Nitzan,
 %\emph{Current induced light and light induced current 
 %      in molecular tunneling junctions},
 Phys. Rev. Lett. {\bf 95}, 206802 (2005);
 %\emph{Optical properties of current-carrying molecular wires},
 J. Chem. Phys. {\bf 124}, 234709 (2006).

\bibitem{Tien63}
 {P. K.} Tien and {J. P.} Gordon,
 %\emph{Multiphoton Process Observed in the Interaction of 
 %      Microwave Fields with the Tunneling between 
 %      Superconductor Films},
 Phys. Rev. {\bf 129}, 647 (1963).

\bibitem{Pedersen98}
 {M. H.} Pedersen and M. B\"uttiker,
 %\emph{Scattering theory of photon-assisted electron transport},
 Phys. Rev. B {\bf 58}, 12993 (1998).

\bibitem{Moskalets04}
 M. Moskalets and M. B\"uttiker,
 %\emph{Adiabatic quantum pump an the presence of external ac voltages},
 Phys. Rev. B {\bf 69}, 205316 (2004).

\bibitem{Kohler04}
 S. Kohler, J. Lehmann, and P. H\"anggi,
 %\emph{Driven quantum transport on the nanoscale},
 Phys. Rep. {\bf 406}, 379 (2005).

\bibitem{Platero04}
 G. Platero and R. Aguado,
 %\emph{Photon-assisted transport in semiconductor nanostructures},
 Phys. Rep. {\bf 395}, 1 (2004).

\bibitem{Tucker85}
 {J. C.} Tucker and {M. J.} Feldman, 
 %\emph{Quantum detection at millimeter wavelengths},
 Rev. Mod. Phys. {\bf 57}, 1055 (1985).

\bibitem{Note1}
 We assume the polarization direction to be along the molecular 
 bridge, since this should give the largest values of $V_{ac}$.

\bibitem{Ahlrichs89}
 R. Ahlrichs, M. B\"ar, M. H\"aser, H. Horn, and C. K\"olmel,
 %\emph{Electronic structure calculations on workstation computers: 
 %The program system turbomole},
 Chem. Phys. Lett. {\bf 162}, 165 (1989).

\bibitem{PaulyThesis}
 F.\ Pauly, Ph.D. thesis, Institut f\"ur Theoretische Festk\"orperphysik,
 Universit\"at Karlsruhe, Karlsruhe (2007).

\bibitem{Wohlthat07}
 S. Wohlthat, F. Pauly, {J. K.} Viljas, {J. C.} Cuevas, and G. Sch\"on,
 %\emph{Charge transport through single oxygen molecules in atomic 
 %aluminum contacts - an ab initio study},
 cond-mat/0702477;  
 F.\ Pauly \emph{et al.}, to be published (2007).

\bibitem{Datta97}
 S. Datta, W. Tian, S. Hong, R. Reifenberger, {J. I.} Henderson, 
 and {C. P.} Kubiak,
 %\emph{Current-Voltage Characteristics of Self-Assembled Monolayers
 %by Scanning Tunneling Microscopy},
 Phys. Rev. Lett. {\bf 79}, 2530 (1997).

\bibitem{Lang00}
 {N. D.} Lang and Ph. Avouris, 
 %\emph{Carbon-Atom Wires: Charge-Transfer Doping, Voltage Drop,
 %and the Effect of Distortions},
 Phys. Rev. Lett. {\bf 84}, 358 (2000).

\bibitem{Mujica00}
 V. Mujica, {A. E.} Roitberg, and {M. A.} Ratner, 
 %\emph{Molecular wire conductance: Electrostatic potential spatial 
 %profile},
 J. Chem. Phys. {\bf 112}, 6834 (2000).

\bibitem{Damle01}
 {P. S.} Damle, {A. W.} Ghosh, and S. Datta,
 %\emph{Unified description of molecular conduction: 
 %From molecules to metallic wires},
 Phys. Rev. B {\bf 64}, 201403(R) (2001).

\bibitem{Nitzan02}
 A. Nitzan, M. Galperin, {G.-L.} Ingold, and H. Grabert,
 %\emph{On the electrostatic potential profile in biased molecular wires},
 J. Chem. Phys. {\bf 117}, 10837 (2002).

\bibitem{Note2}
We only show results for small enough $\alpha$ and
$\omega$, such that the associated electric fields remain of
reasonable magnitude. The maximum values for $V_{ac}/D$ for the
results in Fig.\ \ref{f.curves} are on the order of $10^9$ V/m.  In
experiments, typical incident powers per area of the focused laser
beam are 10$^7$--10$^8$ W/m$^2$ (Refs.\
\onlinecite{Guhr06,Platero04}), which correspond to considerably
smaller fields of $10^5$ V/m. However, due to the field-enhancement
effects mentioned above, the field strengths described by $\alpha$ can
exceed the externally applied fields by orders of
magnitude.\cite{Grafstrom02}

\bibitem{Wold02}
 {D. J.} Wold, R. Haag, {M. A.} Rampi, and C. D. Frisbie,
 %\emph{Distance dependence of electron tunneling through 
 %self-assembled monolayers measured by conducting probe atomic 
 %force microscopy: Unsaturated versus saturated molecular junctions}
 J. Phys. Chem. B {\bf 106}, 2813 (2002).

\bibitem{Kondo04}
 M. Kondo, T. Tada, K. Yoshizawa,
 %\emph{Wire-length dependence of the conductance of oligo(p-phenylene) 
 %dithiolate wires: A consideration from molecular orbitals}
 J. Phys. Chem. A {\bf 108}, 9143 (2004).

\bibitem{Kurth05}
 S. Kurth, G. Stefanucci, {C.-O.} Almbladh, A. Rubio, and {E. K. U.} Gross,
 %\emph{Time-dependent quantum transport: A practical scheme using
 %density-functional theory}
 Phys. Rev. B {\bf 72}, 035308 (2005).

\bibitem{Arrachea06}
 L. Arrachea and M. Moskalets, 
 %\textbf{Relation between scattering-matrix and Keldysh formalisms 
 %for quantum transport driven by time-periodic fields},
 Phys. Rev. B {\bf 74}, 245322 (2006).

\bibitem{Wu05}
 J.\ Wu, B. Wang, J. Wang and H. Guo,
 %\emph{Giant enhancement of dynamic conductance in molecular devices},
 Phys. Rev. B {\bf 72}, 195324 (2005).

\end{thebibliography}
\end{document}